\documentclass{elsart1p}
\usepackage{graphicx}
\usepackage{amssymb}
\begin{document}
\begin{frontmatter}
%
%
%
\title{Test of Symmetries with Neutrons and Nuclei}
%
%
\author{Stephan Paul}
\address{Physik Department E18 - TU-M\"{u}nchen, 85748 Garching, Germany}
\begin{abstract}
Precision experiments at low energies probing weak interaction are a very promising and complementary tool for investigating the structure of the electro-weak sector of the standard model, and for searching for new phenomena revealing signs for an underlaying new symmetry. With the advent of new technologies in particle trapping and production of beams for exotic nuclei as well as ultracold neutrons, we expect one or two orders of magnitude gain in precision. This corresponds to the progress expected by new high luminosity B-factories or the LHC. Domains studied are $\beta$-decays where decay correlations, partial or total decay rates may reveal the nature of the left-right structure of the interaction and the investigation of discrete symmetries. Here the search for a finite electric dipole moment which, due to its CP-violating nature were sensational by itself, could shed light on the structure of the vacuum at very small distances. Last but not least ideas of a mirror world can be extended to the sector of baryons which can be studied with neutrons.
\end{abstract}
\end{frontmatter}

\section{Introduction}
Neutrons and nuclei, though very similar with respect to the microscopic description of weak interaction, offer very different aspects of it which are addressed by the experiments. These in turn are sensitive also to more macroscopic effects (strong interaction - QCD, effects of composite systems). The simplicity of the free neutron offers a rather clear interpretation of results without much of the theoretical corrections as exhibited by composite systems. The characteristics are:
small radiative corrections,
small influence of isospin breaking effects,
no Schiff moments influencing the search for an electric dipole moment (EDM),
but difficulties in handling purely electrically neutral systems.
Nuclei in turn offer:
flexible mixture of Fermi- and Gamow-Teller transitions,
high statistics,
application of cold atom techniques,
high precision mass measurements,
well focusable systems, requiring only small trapping volumes ('easier' magnetic shielding for EDM experiments),
 but sensitivity to nuclear matrix elements and
ambiguous interpretation for EDM experiments.

\par In the course of this article we will outline the strength of each system using particular examples, thereby focussing on new results and planned experiments.
\section{$\beta$-decays of neutrons and nuclei}
The differential $\beta$-decay rate of a neutron can be written as a sum of contributions involving different kinematical variables like momenta, angular momenta and spin of mother and/or daughter particles \cite{jackson1957}:
\begin{equation}\label{beta_decay_rate}
    w\propto 1+
    a_{\beta\nu}\frac{\vec{p_e} \vec{p_{\nu}}}{E_e E_{\nu}}+
    b\frac{m_e}{E_e}+
    \frac{<\vec{J_A}>}{j_A}[A\frac{\vec{p_e}}{E_e}+ B\frac{\vec{p_{\nu}}}{E_{\nu}}+
    D\frac{\vec{p_e}\times \vec{p_{\nu}}}{E_e E_{\nu}}]+
    c[...]
\end{equation}
where the parameters a, b, c, A, B, D depend on ten coupling constants $\emph C_i$, $\emph C_i^\prime$ (i=1..5). The functional relation of this dependence in turn depends on the transition type and thus on $M_F$, $M_{GT}$, $J$, $J^\prime$. This offers sensitivity to different interaction types like scalar (S), pseudo-scalar (P), vector (V), axial vector (A) or tensor (T) and can be studied in neutron or nuclear $\beta$-decay. The standard model assumes pure V-A structure motivated by experiments and upper limits on scalar or tensor interaction are presently at a level of \cite{severijns2006}:
\begin{equation}\label{scalar_tensor}
    -0.067 < \frac{C_S}{C_V} < 0.067 ~;~~
    -0.081 < \frac{C_T}{C_A} < 0.081
\end{equation}
The decay parameters besides of being pure numbers also relate to the underlying discrete symmetries obeyed or violated in the decay. In the latter case a finite value is required for these parameters. \emph{A, B, C} (see eqn. \ref{C-parameter})\emph{, R} are related to parity violation P, while \emph{D} and \emph{R} (see section \ref{time_reversal}) connect to time-reversal violation T. If T invariance holds, all coupling constants \emph{$C_i$} and \emph{$C_i^\prime$} are real. In addition, the decay parameters a and A are related to the axial and vector coupling constants $g_A$ and $g_V$, respectively, via :
\begin{equation}\label{lambda}
    \lambda=\frac{g_A}{g_V}
\end{equation}

\par
It is important to note, that the neutron offers the possibility of experimentally studying all different decay correlations and clean matrix elements involving strong interaction. Nuclei in turn are mostly used to study full decay rates and beta asymmetries, offering different combinations of the underlying coupling constants \emph{$C_i$}, \emph{$C_i^\prime$} due to the large variety of initial and final states available, with variable $J$, $J^\prime$ or pure $M_F$ and $M_{GT}$ transitions. Thus, the two systems are complementary and have all reasons to be studied with highest precision.

\section{$\beta$-Decay Measurements}
\subsection{$\beta-\nu$-correlations}
\begin{figure}[t]
\begin{minipage}[t]{0.475\textwidth}
\centering
\includegraphics[height=40mm]{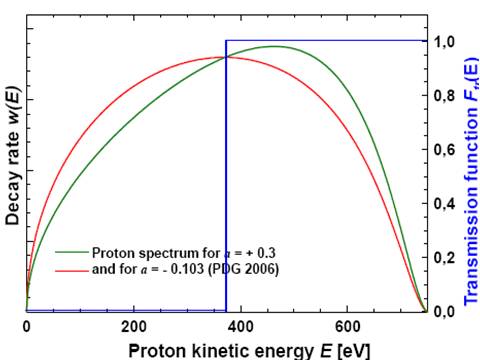}\\
  \caption{Spectral shape of the proton energy from neutron $\beta$-decay for different assumed values for the correlation parameter $a_{\beta\nu}$}\label{a_asymmetry}
\end{minipage}
\hfill
\begin{minipage}[t]{0.475\textwidth}
\centering
\includegraphics[height=40mm]{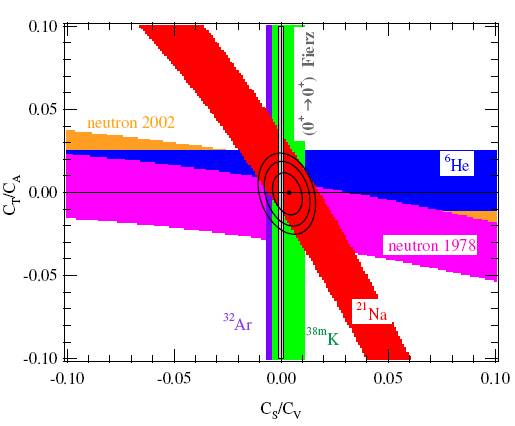}\\
  \caption{Constraints for the scalar and tensor coupling constants $C_S$ and $C_T$ obtained from different measurements \cite{Vetter2007}, assuming couplings to leptons with standard model chirality, see \cite{severijns2006}.}\label{scalar_tensor_coupling}
\end{minipage}
\end{figure}

One decay quantity measured in both, nuclear and neutron decay with high precision is the angular correlation of $\beta$-particle and $\overline{\nu}$. As the $\nu$ cannot be detected, this requires a careful spectroscopical measurement of the recoil nucleus or proton. The momentum distribution of the emerging proton is dependent on the decay (Fermi vs. Gamow-Teller), and the sensitivity of the energy spectrum to the decay parameter $a_{\beta\nu}$ in neutron decay is depicted in figure \ref{a_asymmetry}.
The results obtained so far \cite{Vetter2007} put stringent constraints on the scalar and tensor couplings $C_S$ and $C_T$, as depicted in figure \ref{scalar_tensor_coupling}.
\par
\begin{figure}[ht]
\begin{minipage}[b]{0.45\textwidth}
\centering
\includegraphics[height=34mm]{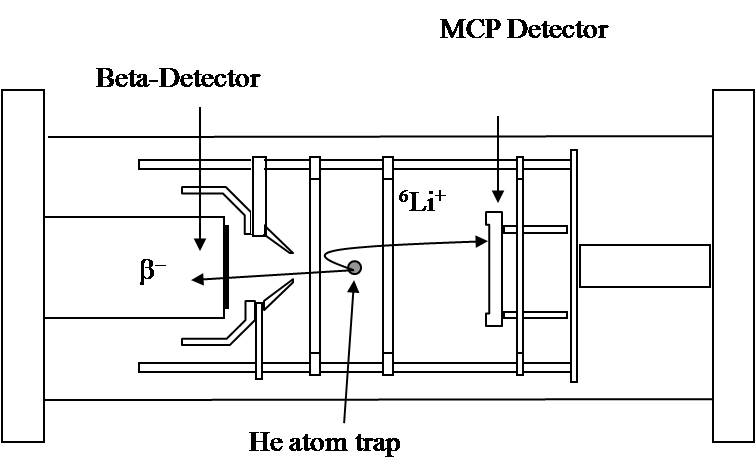}
\caption{Measurement cell for a $\beta$-correlation experiment with nuclei. Electron and recoil nucleus detection are sketched.}\label{nuclear-recoil-trap}
\end{minipage}
\hspace{0.5cm}
\begin{minipage}[b]{0.5\textwidth}
 The decay products are detected in coincidence such that the signal from the $\beta$-particle serves as a start signal for a time-of-flight (TOF) measurement of the proton, detected e.g. in an MCP. Different groups thereby use different isotopes with different transition types. The general accuracy aimed at is $\delta a/a\approx 10^{-3}$. The CAEN-group \cite{flechard2007} studies pure Gamow-Teller transitions in $^6He$ with an accuracy 0.5\%. In turn, pure Fermi-transitions are studied at ISOLDE at

\end{minipage}
\end{figure}

\noindent
In the field of {\bf nuclear}-decays, radioactive isotopes are produced in an accelerator facility like a cyclotron or an
 ISOL-facility. Nuclei are decelerated using techniques of laser cooling \cite{Phillips1998} with
 subsequent trapping in magneto-optical traps (MOT) or dipole traps \cite{Grimm1999}. Other techniques include gas cooling and subsequent
 trapping in Paul- or Penning-traps. A typical set-up is sketched in fig.\ref{nuclear-recoil-trap}.

\par\noindent  CERN using $^{35}Ar$ (with small GT admixture of about 10\%) and in Tri$\mu$p with $^{21}Na$. Presently, the best limit on $\tilde{a}= a/(1 + b\cdot m_e/\langle E_e \rangle) = 0.9981\pm 0.0030 ^{+0.0032} _{-0.0037}$ has been obtained with $^{38}K^m$ \cite{Gorelov2005}.
\par
\begin{figure}[here]
\begin{minipage}[b]{0.45\textwidth}
\centering
\includegraphics[height=44mm]{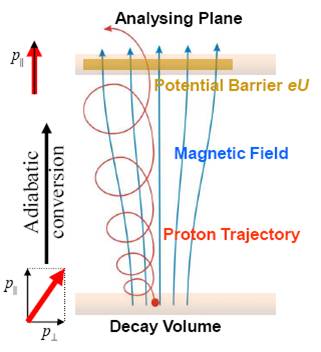}\\
  \caption{Principle of a retardation spectrometer showing magnetic field lines and the particle trajectory}\label{retardation_method}
\end{minipage}
\hspace{0.2cm}
\begin{minipage}[b]{0.54\textwidth}
\par\noindent
Highly polarized ($\wp_n>99\%$) \textbf{neutrons} pass through the central part of a magnetic spectrometer which defines the decay volume. Decay electrons are measured symmetrically at both ends, extracted by a magnetic guiding field. The new PERKEO III experiment \cite{Maerkisch2008} currently been set up is depicted in figure \ref{PERKEO}. The improvement of the statistical accuracy by its predecessor PERKEO II \cite{Reich2000} leads to a value of $\delta A = 0.0007$ \cite{Abele2002} (down from $0.0012$ \cite{Abele1997}). This experiment also measures the proton asymmetry C = 0.2377(26) \cite{Schumann2008_C}, which is related to decay parameters as

\begin{equation}\label{C-parameter}
    N_p\propto \frac{\langle \vec{J_A}\rangle}{J_A} [C \frac{\vec{p_p}}{E_p}]
\end{equation}
\end{minipage}
\end{figure}

For the \textbf{neutron} these investigations use a polarized cold neutron beam. The proton spectrum is measured in a retardation spectrometer, where the transverse motion of protons is fully transformed into a longitudinal one subsequently analyzed by a variable electric field (see fig. \ref{retardation_method}) \cite{simson2008}. Presently the precision is about 5\% with room for further improvement from the present data.

\subsection{$\beta$-decay asymmetry}
The $\beta$ emission from a polarized neutron is described by

\begin{equation}\label{A-parameter}
    N_e\propto b\cdot\frac{m_e}{E_e}+\frac{\langle \vec{J_A}\rangle}{J_A}[A\frac{\vec{p_e}}{E_e}]
\end{equation}

The Leuven group measures the $\beta$-asymmetry using polarized \textbf{nuclei}. They select pure Gamow-Teller transitions but the experiment is done with solid targets which, due to its high mass, causes rescattering of the $\beta$-particles constituing a systematic effect. This limits the accuracy of the experiment to about $\delta A/A = 0.018$, which seems characteristic of the experiments with polarized nuclei.

 \begin{figure}[ht]\begin{centering}
  \includegraphics[width=9.5cm]{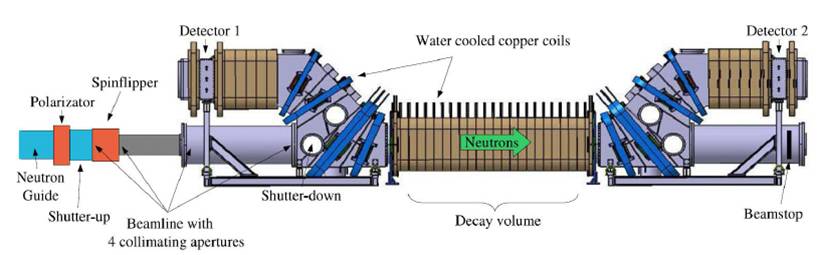}\\
  \caption{The newly installed PERKEO III spectrometer currently operated at ILL \cite{Maerkisch2008}.}\label{PERKEO}
\end{centering}\end{figure}

\subsection{Time-reversal tests}
\label{time_reversal}
Decay correlations can also be used to perform tests for time-reversal violation by measuring the helicity of the $\beta$-particle emerging from a polarized neutron.

 \begin{equation}\nonumber
    N_e(\vec{J_A},\vec{\sigma_e})\propto 1 + \frac{\langle \vec{J_A}\rangle}{J_A}[A~\frac{\vec{p_e}}{E_e}
    + R~\frac{\vec{p_e}\times\vec{\sigma_e}}{E_e}]~,~~\textrm{where}~
 \end{equation}
 \begin{equation}\nonumber\raggedright
 \nonumber R=0.28~ Im(\frac{C_S+C_S^{\prime}}{C_A})+ 0.33~Im(\frac{C_T+C_T^{\prime}}{C_A})
    \label{R-parameter}
\end{equation}

\begin{figure}[t]
\begin{minipage}[t]{0.475\textwidth}
\centering
\includegraphics[height=40mm]{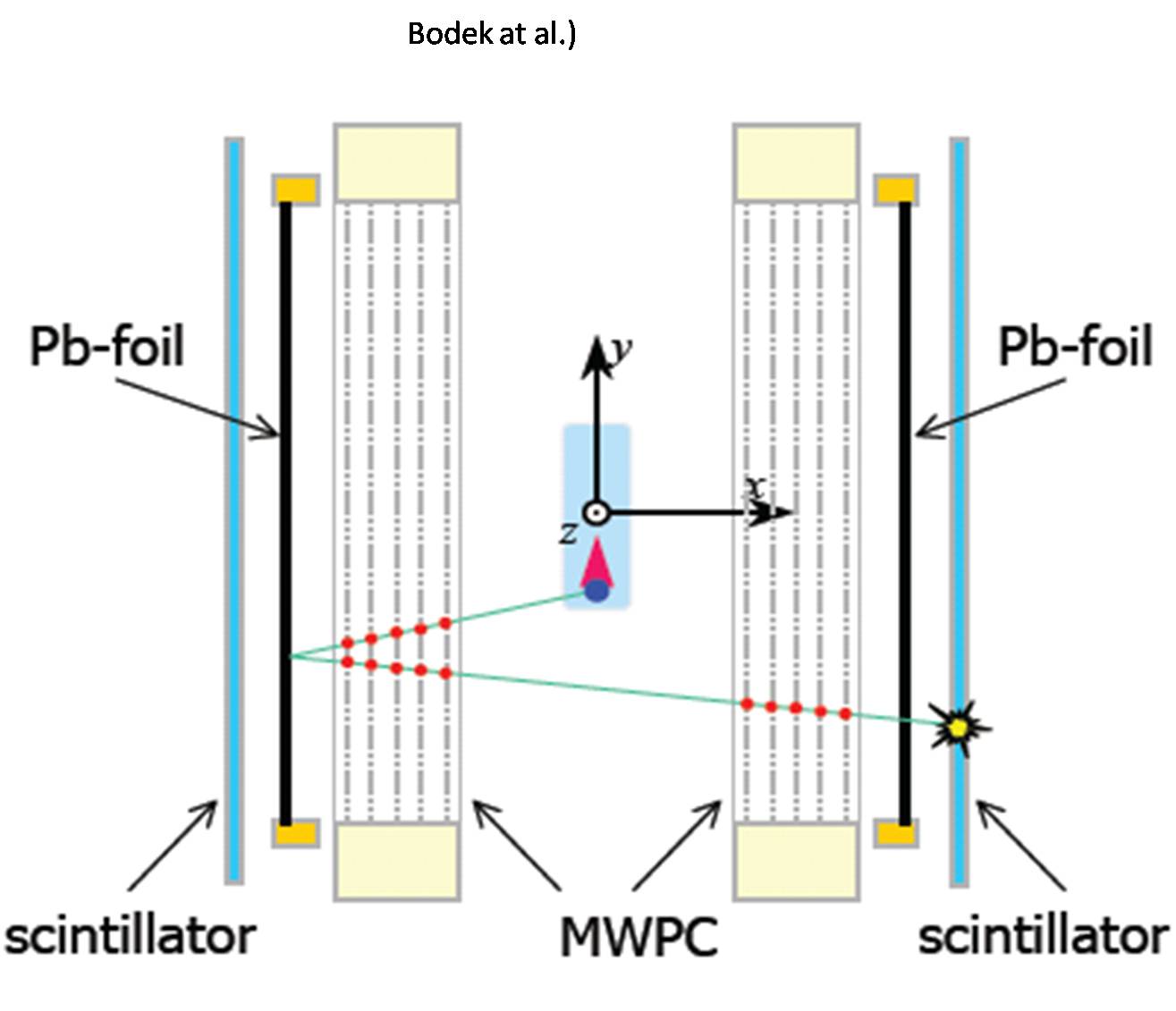}\\
  \caption{The experiment built to measure the $e^-$helicity in neutron $\beta$-decay}\label{time_reversal_exp}
\end{minipage}
\hfill
\begin{minipage}[t]{0.475\textwidth}
\centering
\includegraphics[height=40mm]{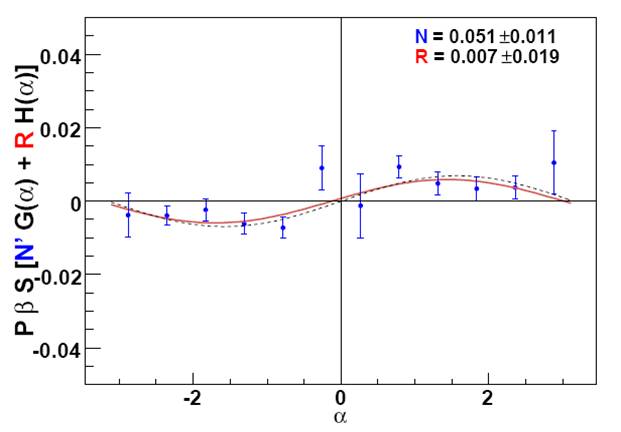}\\
  \caption{Experimental asymmetry versus $\alpha$, the angle enclosed by decay electron and scattering plane, allowing for finite values of the decay parameters N (not introduced here) and R \cite{Bodek2008}.}\label{R-results}
\end{minipage}
\end{figure}

This very difficult triple correlation experiment has been performed with a polarized cold neutron beam at PSI (fig. \ref{time_reversal_exp}), using a Mott-scattering spectrometer to exploit the spin-dependent e-backscattering on a carbon foil. The results are depicted in fig. \ref{R-results} limiting $R$ to $R<0.032$ ($2\sigma$ limit), with potential for yet another factor of 2 improvement \cite{Bodek2008}.

\section{The Hunt for Electric Dipole Moments}
Electric dipole moments (EDM) are in the focus of many research groups as they maybe a key to the understanding of baryogenesis. The existence of an EDM ($d_n$) implies violation of T-symmetry (and thus CP-symmetry) in a flavour diagonal system, which has not been observed so far:
\begin{equation}\label{EDM}
    H = -(d_n \vec{E}\cdot \vec{J} + \mu_n \vec{B}\cdot \vec{J})/|\vec{J}|
\end{equation}

 Neutrons and nuclei are studied with extreme precision, mostly using Ramsey's technique of separated oscillatory fields, where the neutrons (nucleus) Larmor-precession frequency due to its magnetic moment $\mu_n$ directed in the direction of its total angular momentum $\vec{J}$ inside a very homogenous and well controlled magnetic field $\vec{B}$, is compared with an external reference clock. The neutron's precession will change under the influence of an electric field $\vec{E}$, if the EDM of the system under study has a finite value (see eqn. \ref{EDM}).

 While the principle is the same for most experiments, the technique of preparation and analyzing the direction of polarization after the Larmor-precession is different for atoms/nuclei and neutrons. The system studied determines the physics: electron-EDM, nucleon-EDM which itself can originate from different physics processes \cite{pospelov2005} (quark-EDM, QCD- $\theta$-term), nucleon-nucleon interaction and nuclear Schiff moments, for which the nuclear effect is transferred to the atom leading to enhancement factors for a nuclear EDM in the atom by a factor 100 \cite{Auerbach2008}).
 \subsection{Searching for a neutron EDM}
 The present limit for the EDM of the neutron is $d_n<2.9\cdot 10^{-26}e\cdot cm$ \cite{Baker2006}. Several new experiments are being prepared, all  aiming at eventually reaching a limit of $10^{-28}e\cdot cm$ (\cite{Harris2000}, \cite{Bodek_EDM2008}, \cite{Golub1994}, \cite{Aleksandrov2007}). The nEDM collaboration \cite{Harris2000} follows a 2-step process. Using a modified existing set-up, they aim at a sensitivity of  a few $10^{-27}e\cdot cm$. With a newly constructed apparatus, requiring a large size magnetic shield and novel magnetometry employing Cs, $^3He$ and xenon-magnetometers, they aim to improve the sensitivity by yet another order of magnitude by 2012.
 \begin{figure}[t]
\begin{minipage}[t]{0.475\textwidth}
\centering
\includegraphics[height=40mm]{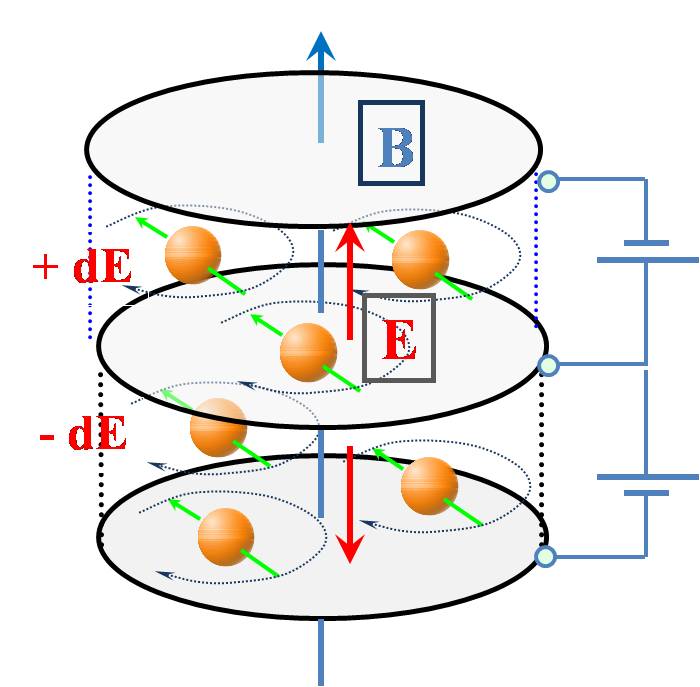}\\
  \caption{Principle of the double chamber system to measure he EDM of the neutron. The two chambers are operated with opposite electric fields but equal direction of the magnetic field. The difference of the Larmor precession frequency $\Delta\omega=\omega_{\uparrow\uparrow}-\omega_{\uparrow\downarrow}=4\cdot d_n\cdot E$}\label{EDM_ramsey}
\end{minipage}
\hfill
\begin{minipage}[t]{0.475\textwidth}
\centering
\includegraphics[height=40mm]{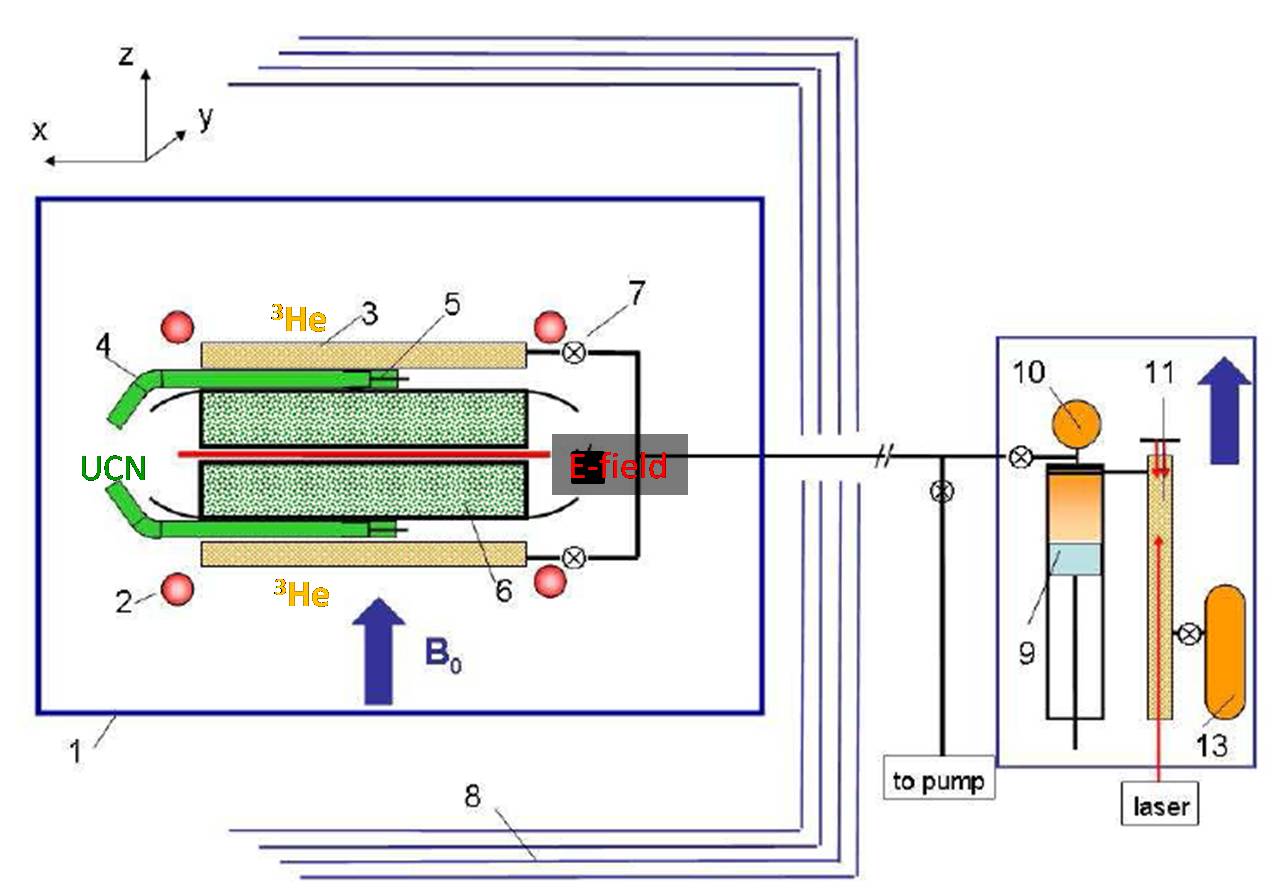}\\
  \caption{Planned set-up of the new joint EDM experiment by the nEDM collaboration planning to work at the new UCN-sources at PSI and Munich. Note the large dimensions of the magnetic shielding of about 5m length.}\label{n_EDM}
\end{minipage}
\end{figure}

 \subsection{EDM searches with nuclei}
\begin{figure}[here]
\begin{minipage}[t]{0.475\textwidth}
\centering
\includegraphics[height=40mm]{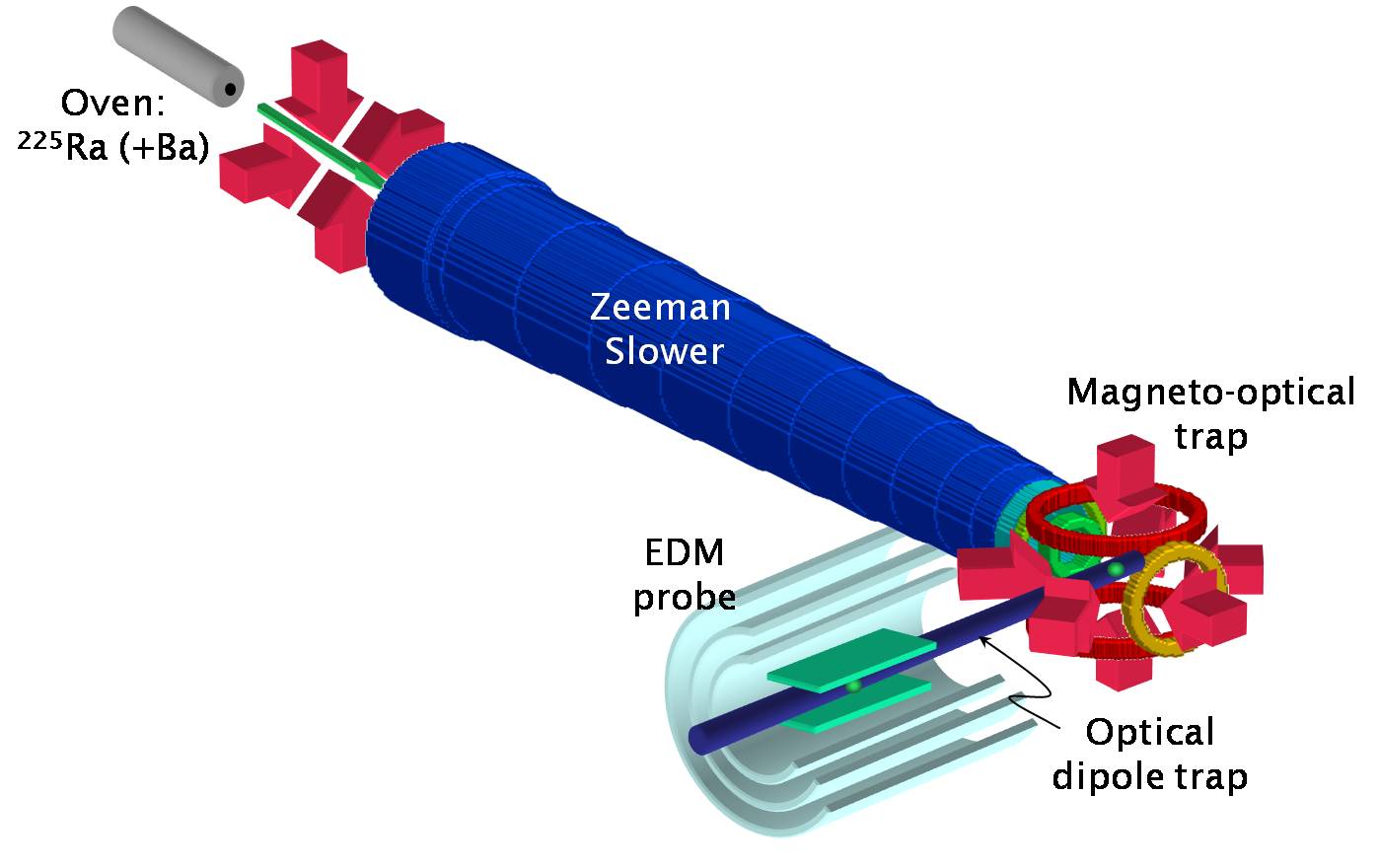}\\
  \caption{Sketch of the set-up at Argonne using a radon beam with subsequent cooling and intermediate storage in a MOT. Also shown is the small Ramsey-cell with magnetic shielding.}\label{EDM_Argonne}
\end{minipage}
\hfill
\begin{minipage}[t]{0.475\textwidth}
\centering
\includegraphics[height=40mm]{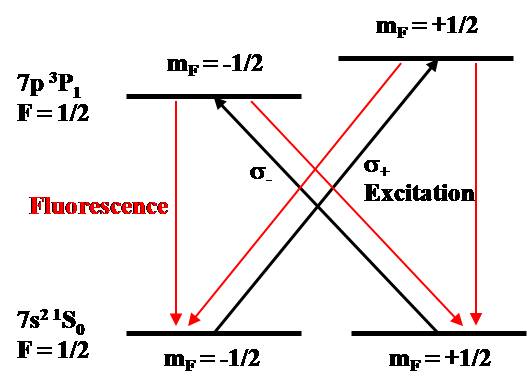}\\
  \caption{Optical level and excitation scheme for the preparation and analysis of polarized nuclei inside the EDM-trap.}\label{nuc_EDM_level}
\end{minipage}
\end{figure}

 The case of nuclear EDM (and/or electron EDM) is different, experimentally. Beams of radioactive nuclei can be cooled and subsequently stored in small size traps making the set-up of the Ramsey-cell much simpler than for neutrons. Here much effort has been spent for proper beam cooling and storage techniques as well as for the preparations of exotic beams. The Argonne group \cite{Guest2007} is preparing a new setup using Radium beams shown in fig. \ref{EDM_Argonne}. The polarization of nuclei and its reading is performed using circular polarized laser beams operating at a pair of Zeeman split levels (fig. \ref{nuc_EDM_level}).
A novel technique has recently been proposed by a group at TU-M\"unchen \cite{Fierlinger2009}, using $^{129}Xe$ as a probe. Small drops of xenon are condensed onto a small chip into shallow structures, each surrounded by a circle of small conductive pillars. The setup is integrated into a homogenous magnetic field. The atomic Larmor-precession is observed by means of small squids place underneath each of the droplets. As a novelty in this field of research, the electric field now is applied to the pillars, perpendicular to the B-field, and is not constant in space but co-rotating around the droplet with the Larmor-precession frequency. Thus, a finite value of an EDM results into a movement of the Xe-spin out of the original plane of Larmor-rotation, building up with time. This set-up is presently being completed and first results are expected end of 2009. The final sensitivity aimed at is $d_{Xe}<10^{-30}e\cdot cm$.

\section{Mirror Neutrons}
The concept of mirror particles is borne out of the idea to restore parity symmetry described by V-A theory by adding the V+A sector to the Langrangian. This requires the existence of mirror particles of the same mass as our known particles and identical couplings between the mirror particles themselves (see e.g. \cite{Berezhiania2006}). The coupling between particle and mirror particle comes through gravity which causes a mixing of the two sectors with oscillation times $\tau_{nn^\prime}\gg \tau_n$. There is no $\Delta B=2$ transition necessary as in the case of $n\leftrightarrow \overline{n}$ oscillations searched for many years ago at ILL \cite{Baldo_Ceolin1994}. The transition probability is given by
\begin{equation}\label{oscillation}
    P_{nn^\prime}(t) = <n^\prime\mid H(t)>^2 = \frac{\sin^2{\sqrt{1+(\omega\tau_{nn^\prime})^2}\cdot t/\tau_{nn^\prime}}}{1+(\omega\tau_{nn^\prime})^2}
\end{equation}

\par
\begin{figure}[h]
\begin{minipage}[t]{0.475\textwidth}
\centering
\includegraphics[height=40mm]{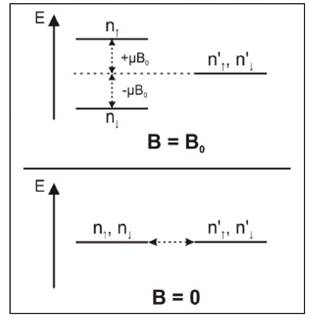}\\
  \caption{Principle of neutron-mirror neutron mass splitting for a magnetic field B with $B=0$ and $B\neq 0$}.\label{oscillation_scheme}
\end{minipage}
\hfill
\begin{minipage}[t]{0.475\textwidth}
\centering
\includegraphics[height=40mm]{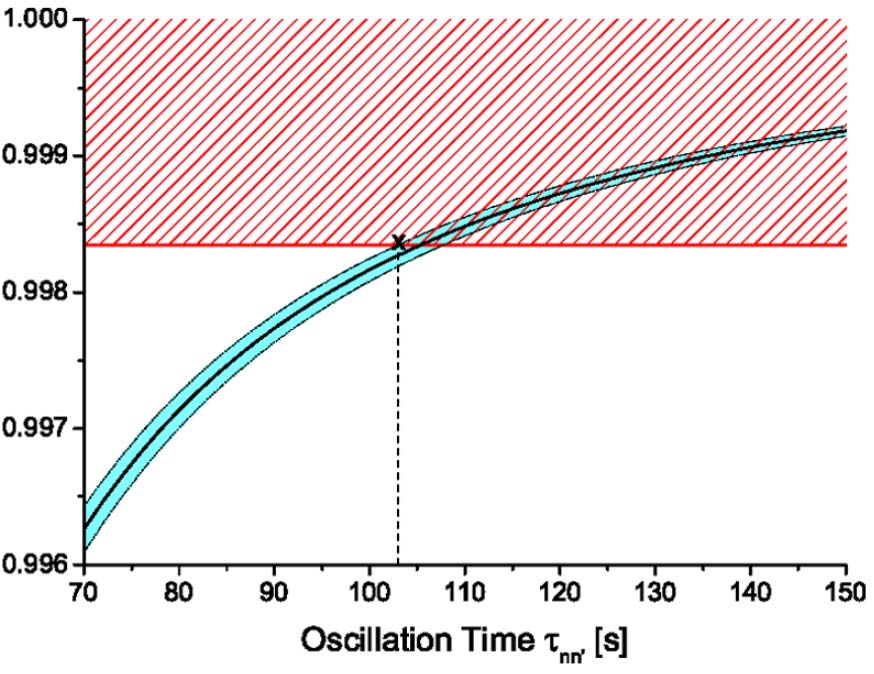}\\
  \caption{Result of the ${nn^\prime}$-oscillation experiment \cite{Ban2007}. The ratio of neutrons observed after storage with and w.o. magnetic field is depicted. }\label{oscillations_result}
\end{minipage}
\end{figure}

One way to search for these transitions is a disappearance experiment comparing the lifetime of neutrons inside a storage volume with and without magnetic field, as a B-field can cause an energy splitting between the two degenerate mass-eigenstates (fig. \ref{oscillation_scheme}), thus strongly damping possible oscillations. Owing to their excellent shielding for external B-fields, EDM-setups are ideally suited for this search. Two groups have recently performed such a search, exploring $P_{nn^\prime}(t)$ for different values of $\omega$. For $B\ll 50nT$ we get $\omega\ll 1$ and $P_{nn^\prime}(t)$ approaches $t^2/\tau_{nn^\prime}^2$. For $B\gg 5\mu T$, $\omega \gg 1$ and $P_{nn^\prime}(t)$ approaches $1/(2\cdot \omega\tau_{nn^\prime})^2$.

\par\noindent The results for the OILL-group at ILL \cite{Ban2007} is depicted in fig. \ref{oscillations_result}. Their upper limit for $\tau_{nn^\prime}$ is $\tau_{nn^\prime} > 103s$ (at $95\%$ CL). Shortly after, an improved lower limit was obtained by \cite{Serebrov2007} with $\tau_{nn^\prime} > 448 s$. These result improve previous measurements by a factor 400.
\par
The physics has direct consequences also for ultra-high energy (UHE) cosmic rays, as mirror particles have longer flight path as fewer $N\gamma$-reactions occur leading to more UHE nucleons than assumed \cite{Berezhiania2006}.

\end{document}